\documentclass{aastex}
\usepackage{spr-astr-addons}
\usepackage{url}

\RequirePackage{color}

\newcommand{\emaila}{mohsen.shadmehri@dcu.ie}

\begin{document}

\title{Dynamics of charged dust particles in protoplanetary discs}
\shorttitle{Dynamics of Dust Particles in Discs}
\shortauthors{M. Shadmehri}

\author{Mohsen Shadmehri\altaffilmark{1,2}}

\email{\emaila}

\altaffiltext{1}{School of Mathematical Sciences, Dublin City University, Glasnevin, Dublin 9, Ireland}
\altaffiltext{2}{Department of Physics, School of Science, Ferdowsi University, Mashhad, Iran}

\begin{abstract}
We study the effect of an imposed  magnetic field on the motion of
charged dust particles in  magnetically active regions of a protoplanetary disc. Assuming a
power law structure for the vertical and the toroidal components of
the magnetic field for the regions beyond magnetically dead region
of the disc, the radial and the vertical velocities of the charged
particles, in the asymptotic case of small particles, are calculated
analytically. While grains with radii smaller than a critical radius
significantly are affected by the magnetic force, motion of the particles
with larger radii is independent of the magnetic field. The critical
radius depends on the magnetic geometry and the charge of the
grains. Assuming that a grain particle has one elementary charge and
the physical properties of the disc correspond to a minimum-mass
solar nebula, we show that only micron-sized grains are affected by
the magnetic force. Also, charge polarity determines direction of
the radial velocity. For such small particles, both the radial
and the vertical velocities increase  due to the magnetic
force.
\end{abstract}



\section{Introduction}
Physical structures of an accretion disk are determined by various
physical factors such as gravity of the central star, turbulence,
magnetic fields. Understanding the time evolution of dust particles
in circumstellar disc-like structures around protostars and young
stellar objects is crucial for explaining the formation of planetary
systems. Due to various complex physical processes that are playing
significant roles in dust and gas discs, it is difficult to present
a complete and self-consistent model for such systems. Because of
the gas drag force, the motion of the dust particles is different
from that of gas (e.g., Weidenschilling 1977). The gas drag force
takes angular momentum from particles, resulting in their inward
migration (Whipple 1972). Garaud, Barri$\grave{e}$re-Fouchet \& Lin
(2004) found that small particles move smoothly to the midplane,
while large bodies oscillate about the midplane with decreasing
amplitude.

Takeuchi \& Lin (2002) (here after TL) studied grains motion in
optically thick disc, using the assumption that radiation pressure
on the dust is negligible. They showed that as the gas accretion
proceeds, small dust particles are left behind in the disc, and the
dust to gas ratio increases. In a subsequent paper, Takeuchi \& Lin
(2003) investigated the outflow of dust particles on the surface
layers of optically thick discs. Klahr \& Lin (2001) studied the
motion of grains in optically thin disc, where the dust experiences
both gas drag and radiation pressure. In a similar analysis,
Takeuchi \& Artymowicz (2001)  suggested that ring-like features
which have been observed in some circumstellar debris discs (e.g.
Schneider et al. 1999; Marsh et al. 2005), arise naturally in discs
passing through the transitional phase from gas-dominated to
dust-dominated discs. The dust particles migrate radially until they
are in co-rotation with the gas.

To our knowledge, most of the previous studies are assuming that the
motion of particles in the gas and dust accretion discs exclusively
caused by gravitation, gas drag force and radiation pressure of the
central protostar. However, collisional dust grain electrification
in the early solar nebula has long been discussed to be a cause for
lightning, which could explain the melting of meteoritic chondrule
precursors. Poppe, Blum $\&$ Henning (2000) studied collisional
grain charging experimentally. They showed that collisional grain
charging is stronger than previously discussed with respect to
preplanetary grains and should be considered concerning the
preplanetary dust aggregation and the formation of lightning in the
solar nebula. It also has influences on other physical conditions in
gas and dust discs, because charged particles can interact with
the magnetic field. Although the strength and direction of magnetic
fields in accretion discs are rather unknown, the observation of
winds and outflows is explained by magnetic acceleration mechanisms
that require the existence of a magnetic field with strength of
$\approx 10^{-4}$ T at 1 AU solar distance (K$\ddot{o}$nigl $\&$
Ruden 1993).

The possible effects of magnetic fields on the structure of
accretion discs have been studied both numerically or analytically
using simplified models (e.g. Lovelace, Romanova \& Newman  1994;
Bisnovatyi-Kogan \& Lovelace 2000; Igumenshchev, Narayan \&
Abramowicz 2003). These studies, generally show that magnetic fields
may have significant effects on the structure of the discs. So, in
gas and dust discs, one may expect such type of behaviors due to
the magnetic fields.

However, the ionization structure of protoplanetary discs which determines the coupling between the magnetic field and the grains is
likely to be inhomogeneous and complex. The ionization fraction at different
locations, and even different times, are not easy to calculate (e.g.
Ilgner \& Nelson 2006). In fact, the idea of a magnetically dead-zone was first proposed by Gammie (1996). He argued that there is a poorly ionized region where the growth of magnetorotational instability against ohmic dissipation can not be sustained. The dead zone typically stretching out to around 10 AU to 15 AU (e.g. Reyes-Ruiz 2001; Matsumura \& Pudritz 2006). Wardle (2007) studied the ionization equilibrium and magnetic diffusivity as a function of height from the disc midplane at radii 1 and 5 AU.  He showed that the dead zone may not extend over the entire vertical dimension of the disc, because not only the ionization fraction but the gas density are  strong functions of height above the midplane and so the magnetic coupling is likely to increase away from the midplane. Moreover, the ionization fraction itself is a strong function of the presence (and size) of the dust particles mixed with the gas and recent calculations have shown that once the grains have settled, the ionisation level is sufficient to couple the magnetic field to the gas at 1 AU, even at the midplane (Wardle 2007). But the disc becomes magnetically active beyond the dead zone (e.g. Reyes-Ruiz 2001). In this study, we are applying a simplified and phenomenological model for describing the gas component of a gas
and dust disc for regions beyond the dead zone.

We are
assuming only the dust particles which are influenced by an imposed
magnetic field.  The properties of any external magnetic fields
threading a dust and gas disc are not well-known. However, it is
very unlikely that the dipolar field of the central protostar is
important at several AU from the protostar where the planets are
believed to form. More probably such an external field comes from
the the molecular cloud core out of which the gas and dust disc
formed.

These considerations show that magnetic forces on micron-sized
particles in gas and dust disks will be worth further investigation.
In fact, the structure of magnetic fields within protostellar discs
may be studied via polarimetry provided that grains are aligned in
respect to magnetic field within the discs. In this study, we assume
that the charged particles are collisionless and indestructible
spheres in a gaseous disc and, to keep the problem tractable, we
also neglect the feedback of the particles' drag on the motion of
the gas.  We mainly follow the approach of TL in analyzing the
dynamics of grains, but in an imposed magnetic field. We show that a
vertical imposed magnetic field affects the radial and the vertical motions of charged
dust particles. The basic equations of the model are described in
section 2. In sections 3 we solve for individual particles'
velocities, which are charged and small with sizes smaller than the mean
free path of the gas molecules. In the final section, we summarize
our results.

\section{General Formulation}

We study the motion of a dust particle with mass $m$ and charge $q$
in a magnetized gas disc. One can simply show that the collisional
timescale is longer than the stopping time by at least an order of
magnitude (Youdin \& Shu 2002). So, we can ignore the effect of
collisions. For a charged dust particle, Newton's second law can be
written as
\begin{equation}
m\frac{d^{2}{\bf r}}{dt^{2}}={\bf F}_{\rm grav}+{\bf F}_{\rm g}+{\bf F}_{\rm mag}+{\bf F}_{\rm electric},
\end{equation}
where $\bf r$ is the radius vector of the particle. Also, ${\bf F}_{\rm grav}$, ${\bf F}_{\rm g}$, ${\bf F}_{\rm
mag}$ and ${\bf F}_{\rm electric}$ are the gravitational force of the central star, the  gas drag, the magnetic
force and the electric force, respectively. We neglect radiation
pressure and Poynting-Robertson drag for simplicity. It is more
convenient to write the equations of motion in cylindrical
coordinates $(r, \theta)$.

As for the magnetic field geometry, we consider the toroidal and the
vertical components of the magnetic field. If a charged grain is
placed in a rotating frame, it will invariably feel an electric
field. The Keplerian velocity is denoted by $v_{\rm
K}=(GM_{\ast}/r)^{1/2}$, where $M_{\ast}$ is
the mass of the central star. We can imagine one of two approaches: (a) either
state  that the equation is written in a frame where electric field
is zero, i.e. ${\bf E} + {v_{\rm K} {\bf e}_{\theta}} \times {\bf B}
= 0$ in which case Lorentz Force term will become $q ({\bf v} -
{v_{\rm K} {\bf e}_{\theta}}) \times {\bf B}$; or (b) explicitly
include an electric field term. We follow the first approach. In highly conducting plasma the electric field in the frame co-moving with the neutrals vanishes, so that the current density remains finite. But if we include the evolution of the magnetic field via the induction equation, this is not a good approximation in the presence of the charged dust grains. Because charged dust particles can easily decouple from the field lines. In order to avoid this difficulty, the magnetic field is treated as an imposed constraint in this study.

Now we can write the radial and the azimuthal equations of motion,
\begin{displaymath}
\frac{dv_{\rm r,d}}{dt}=\frac{v_{\theta,d}^{2}}{r}-\frac{v_{\rm
K}^{2}}{r}-\frac{v_{\rm K}}{T_{\rm s}r}(v_{\rm r,d}-v_{\rm
r,g})
\end{displaymath}
\begin{equation}
+\frac{q}{m} [(v_{\theta,d}-v_{\rm K}) B_{\rm z} - v_{\rm z}
B_{\theta}],\label{eq:main1}
\end{equation}
\begin{equation}
\frac{d}{dt}(rv_{\theta,d})=-\frac{v_{\rm K}}{T_{\rm
s}}(v_{\theta,d}-v_{\rm\theta, g })-\frac{q}{m}rv_{\rm r,d} B_{\rm
z},\label{eq:main2}
\end{equation}
where $T_{\rm s}$ is the nondimensional stopping time. Also, $v_{\rm
r,d}$ and $v_{\theta,d}$ are the $r -$ and $\theta -$ components of
the velocity of the dust particle, i.e. $v_{\rm r,d}=dr/dt=\dot{r}$
and $v_{\theta,d}=r d\theta/dt= r\dot{\theta}$. Note that
$v_{\rm\theta, g}$ is the rotational velocity of the gas component.

In order to write the $z-$component of the equation of motion, we
assume the particles reach terminal velocity, at which the gravity,
the gas drag and the Lorentz force balance with each other,
\begin{equation}
\frac{m v_{\rm z,d} \Omega_{\rm K}}{T_{\rm s}}=-m\Omega_{\rm
K}^{2}z+ q v_{\rm r,d} B_{\rm\theta},\label{eq:main3}
\end{equation}
where $\Omega_{\rm K}=v_{\rm K}/r$.

The nondimensional stopping time $T_{\rm s}$ can the be written as
(TL),
\begin{equation}
T_{\rm s}=\frac{\rho_{\rm d} s v_{\rm K}}{\rho_{\rm g} r v_{\rm
T }},\label{eq:Tss}
\end{equation}
where $v_{\rm T}$ represents the $\sqrt{8/\pi}$ times the mean thermal
velocity $c$ of the gas. Also, $\rho_{\rm d}=1.25$ g
cm$^{3}$ is the grain bulk density and $\rho_{\rm g}$ is the gas density. In writing the above relation a
spherical shape with radius $s$ is assumed for the grain.

We now specify our ansatz for the magnetic field. The vertical component of the 
field is assumed to take a form
\begin{equation}
B_{\rm z}=B_{0}(\frac{R}{r})^{n},
\end{equation}
where $R=1$ AU $=1.5\times 10^{13}$ cm and $B_{0} = 10^{-4}$ T is
the strength of the field at this radius. The normalization for the
distance occurs relative to 1 AU. We will also consider more general
values of $n$ in order to explore the input parameter space. Such a
power-law relation for the vertical component of the magnetic field
has been studied theoretically in self-similar models (e.g.,
Shadmehri \& Khajenabi 2006; Wu \& Lou 2006). For example, a
distributed $B_{\rm z}$ field with $n=5/4$ has been suggested by
Blandford \& Payne (1982). The paraboloidal field model of Blandford
(1976) corresponds to $n=1$, and the models discussed by Ostriker
(1997) corresponds to the range $0<n<1$. Also, the toroidal
component of the magnetic field is (Lovelace, Romanova \& Newman
1994)
\begin{equation}
B_{\theta}=\frac{z}{h_{\rm g}} B_{\theta,S},\label{eq:Btheta}
\end{equation}
where $B_{\theta,S}=\beta_{\theta} B_{\rm z}$ and $\beta_{\theta}$
is constant of order unity ($\beta_{\theta} < 0$).

The other physical variables of the gas component are
\begin{equation}
\rho_{\rm g}(r, z)=\rho_{0} (\frac{r}{R})^{p} \exp(-\frac{z^2}{2
h_{\rm g}^2}),
\end{equation}
\begin{equation}
c(r)=c_{0} (\frac{r}{R})^{q/2},
\end{equation}
\begin{equation}
h_{\rm g}(r)=h_{0} (\frac{r}{R})^{(q+3)/2},
\end{equation}
where the power-law indices $p$ and $q$ are generally negative.
Also, the radial and the rotational velocities of the gas are
\begin{displaymath}
v_{\rm r,g}=-2\pi\alpha (\frac{h_0}{R})^{2}
(3p+2q+6+\frac{5q+9}{2}\frac{z^2}{h_{\rm g}^2})
\end{displaymath}
\begin{equation}
\times (\frac{r}{R})^{q+1/2} \rm AU yr^{-1},
\end{equation}
\begin{equation}
\Omega_{\rm g} = \Omega_{\rm K} (1-\eta)^{1/2},
\end{equation}
where $\alpha$ is  coefficient for the turbulent viscosity (Shakura
\& Sunyaev 1973), and $\eta$ is written as
\begin{equation}
\eta=-(\frac{h_{\rm g}}{r})^{2} (p+q+\frac{q+3}{2}\frac{z^2}{h_{\rm
g}^{2}}).
\end{equation}

For the overall disc properties, it is often sufficient to work with a disc model in which dust and gas are well mixed and coupled. In our analysis, we adopt the values correspond to a minimum-mass
solar nebula: $M_{\ast}= 1 M_{\odot}$, $\rho_{0}=2.83\times
10^{-10}$ g cm$^{-3}$, $h_{0}=3.33\times 10^{-2}$ AU, $p=-2.25$,
$q=-0.5$ and $\alpha=10^{-3}$. Having these values the gas mass
inside $100$ AU is $2.5\times 10^{-2}$ $M_{\odot}$.

\section{Analysis}

We can obtain approximate analytical solutions for the equations of
the motion by applying some simplifying assumptions. From equations
(\ref{eq:main3}) and (\ref{eq:Btheta}), we obtain
\begin{equation}
v_{\rm z,d}=-T_{\rm s} [\Omega_{\rm K} + \beta_{\theta}
\frac{\Omega}{\Omega_{\rm K}} \frac{1}{h_{\rm g}}\frac{B_{\rm
z}}{B_{0}} v_{\rm r,d}] z,\label{eq:vz1}
\end{equation}
where $\Omega=(-q/m)B_{0}$ is, in fact, the {\it Larmor frequency}
which is defined for describing the motion of a collection of moving
charged particles in a region of constant magnetic field.

We assume that rotational profiles of the gas and the dust components
are close to Keplerian, i.e., $v_{\rm\theta,g}\approx
v_{\rm\theta,d}\approx v_{\rm K}$. Thus, $d(r v_{\rm\theta,d})/dt
\approx v_{\rm r,d} d(r v_{\rm K})dr\approx v_{\rm r,d} v_{\rm
K}/2$. Now, we can rewrite equation (\ref{eq:main2}) as
\begin{equation}
v_{\rm\theta,d}-v_{\rm\theta,g}=v_{\rm r,d}T_{\rm
s}[\frac{\Omega}{\Omega_{\rm K}} (\frac{R}{r})^{n}-\frac{1}{2}].
\end{equation}

In equation (\ref{eq:main1}), we can neglect the left hand side of
the equation if $(v_{\rm r,d}/v_{\rm K})^{2} \ll 1$. Thus, equation
(\ref{eq:main1}) can be written as
\begin{displaymath}
-\eta v_{\rm K}^{2} + 2 v_{\rm K}
(v_{\rm\theta,d}-v_{\rm\theta,g})-\frac{v_{\rm K}}{T_{\rm s}}(v_{\rm
r,d}-v_{\rm r,g})
\end{displaymath}
\begin{equation}
+r\frac{q}{m} [(v_{\theta,d}-v_{\rm K}) B_{\rm z} -
v_{\rm z} B_{\theta}] = 0.
\end{equation}
Substituting  from equation (\ref{eq:vz1}) for $v_{\rm z,d}$ into
the above equation, an algebraic equation for $v_{\rm r,d}$ is
obtained and so,
\begin{equation}
v_{\rm r,d}=\frac{T_{\rm s}^{-1} v_{\rm r,g} + \Gamma_{1} v_{\rm
K}}{\Gamma_{2} T_{\rm s} + T_{\rm s}^{-1}},\label{eq:vr}
\end{equation}
where
\begin{equation}
\Gamma_{1}=-\eta-\beta_{\theta} \frac{\Omega}{\Omega_{\rm
K}}\frac{h_{\rm g}}{r}(\frac{z}{h_{\rm g}})^{2} (\frac{R}{r})^{n}
T_{\rm s},
\end{equation}
\begin{equation}
\Gamma_{2}=1-2\frac{\Omega}{\Omega_{\rm K}} (\frac{R}{r})^{n} +
\beta_{\theta}^{2} (\frac{\Omega}{\Omega_{\rm K}})^{2}
(\frac{z}{h_{\rm g}})^{2} (\frac{R}{r})^{2n}.
\end{equation}
\begin{figure}
\plotone{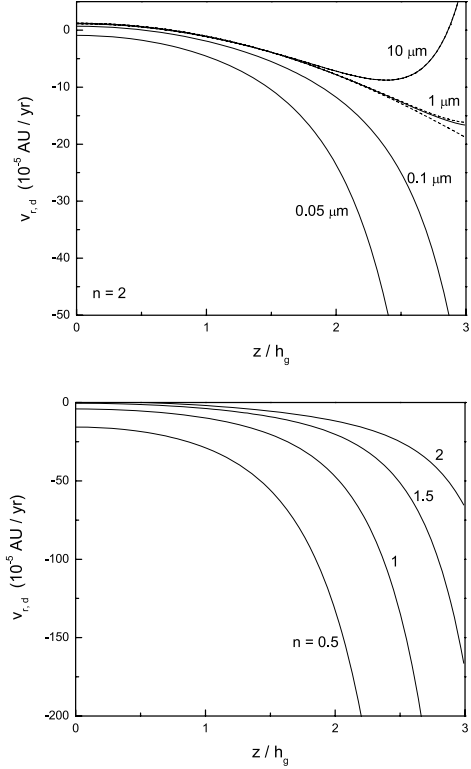}
\caption{({\it top}) Radial velocity $v_{r,d}$ of dust particles  of
$s=0.05$ $\mu$m, $0.1$ $\mu$m, $1$ $\mu$m and $10$ $\mu$m at the
radial distance $10$ AU from the central object. The distance from
the midplane $z$ is normalized by the disc scale hight $h_{g}$.
Dashed curves show profiles of the nonmagnetic solutions. The
exponent of the magnetic field is $n=2$. The electrical charge of a
particle is $q=-e$, where $e$ is the elementary charge. For
particles with $s=0.1$ $\mu$m the difference between the magnetic
(solid curve) and the nonmagnetic (dashed curve) solutions is
negligible. Profiles of the magnetic and the nonmagnetic solutions are the same for particles with larger radii (e.g. $s=10$ $\mu$m).
But for smaller particles (e.g. $s=0.1$ and $0.05$ $\mu$m) the
profiles significantly deviate from the dashed curve which shows
nonmagnetic solutions for the same particles. ({\it bottom}) Radial
velocity $v_{r,d}$ of dust particles of $s=0.1$ $\mu$m for various
exponent of the magnetic field $n=0.5$, $1$, $1.5$ and $2$. Here,
the dust particles are negatively charged with $q=-e$. }
\label{fig:1}
\end{figure}

We can study profile of the radial velocity of dust particles using equation (\ref{eq:vr}). If we substitute this relation into equation (\ref{eq:vz1}), an analytical solution is obtained for the vertical velocity of the dust particles. Actually, for neutral dust particles, both parameters $\Gamma_1$ and $\Gamma_2$ become one and our analytical equations (\ref{eq:vz1}) and (\ref{eq:vr}) reduce to the solutions of TL. The effect of the magnetic field on charged particles appears through Larmor frequency $\Omega$ and the parameter $\beta_{\theta}$. However, in our proposed geometry of the magnetic field, we have an order of unity value for $\beta_{\theta}$. So, dependence of the solutions on this parameter is weak and in our subsequent analysis we assume $\beta_{\theta} \approx -1$. But the Larmor frequency $\Omega$ not only depends on the magnetic strength $B_0$, but mass and the electrical charge of a dust particle are important. We consider micron-sized grains where the nondimensional stopping time is much smaller than unity through most of the disc. Such small particles are well coupled to the gas component.  Poppe, Blum \& Henning (2000) demonstrated  experimentally that a micron-sized grains may
acquire 100 elementary charge. However, in cold protoplanetary
discs, grains are either neutral or have 1 to 2 electronic charges. It is very unlikely
planetary disc model where grains can carry 100 charges near the
mid-plane. It may be possible in the hot HII regions where
temperature is very high, but in our analysis we consider 1 electrical charge. However, this situation changes significantly  as we consider higher vertical locations. For example, for $r = 1$  AU, the mean grain charge is $-14e$ above $3 h_{\rm g}$ to $4h_{\rm g}$  (Wardle 2007). At larger radii, this transition may occur even closer to the midplane, given the drop in fluid density away from the central object. Since our solutions  reach to $z/h_{\rm g} = 3$, it is plausible that at those locations the typical grain carries more than a single charge.

\begin{figure}
\plotone{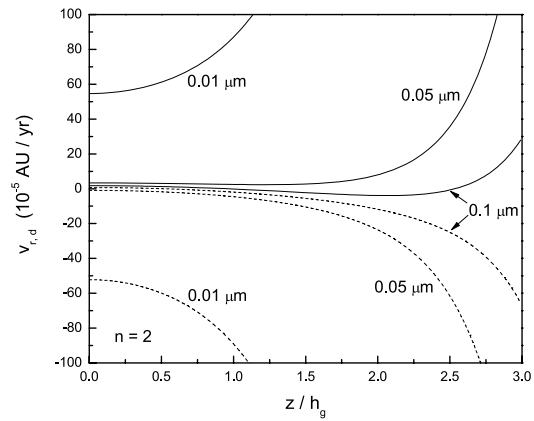}
\caption{The same as Figure \ref{fig:1}, but for positively charged particles with $q=+e$ (solid curves) and the negatively charged particles with $q=-e$ (dashed curves).} \label{fig:2}
\end{figure}

We are interested in regions of the disc where magnetically are active. As we discussed such regions must be beyond dead region which may extend up to $10-15$ AU. Although we consider the radial distance $10$ AU in our Figures as starting radius of active region, we may simply consider larger radial distances in our analytical solutions. Figure \ref{fig:1} shows radial velocity $v_{\rm r,d}$ versus vertical distance $z$ from the midplane of the disc for different dust sizes (top plot) and various exponent $n$ of the magnetic field (bottom plot). In both plots the electrical charge of a dust particle is $q=-e$, where $e$ is magnitude of the charge of the electron. In the top plot of Figure \ref{fig:1}, we fix the exponent $n=2$. Also, solid curves correspond to the magnetic solutions and the dashed curves represent nonmagnetic solutions. For a particle with radius $1$ $\mu$m, the difference between magnetic and the nonmagnetic solutions is negligible. Also, as the radius increases, the effect of the magnetic field becomes weaker so that for a particle with radius $10$ $\mu$m, magnetic and nonmagnetic solutions are indistinguishable. But for particles with radii smaller than $1$ $\mu$m, the radial profile of the nonmagnetic solution is represented by a dashed curve which is actually radial profile of the gas component as well. For such small particles, magnetic forces significantly modify profile of the radial velocity. In fact, the radial velocity of charged particles increases and as a particle becomes smaller, the radial velocity increases too. Note to the minus sign of the radial velocity. In other words, the radial migration of charged dust particles with radii smaller than critical $s_{\rm c}=1$ $\mu$m significantly increases due to the magnetic force.

Bottom plot of Figure \ref{fig:1} shows radial velocity of a particle with radius $0.1$ $\mu$m for various exponent $n$. The radial profile highly depends on the radial profile of the magnetic field. As the exponent $n$ decreases, the radial migration of a charged grain increases. It implies that the critical radius $s_{\rm c}$ depends on the exponent $n$ as well. While for $n=2$ the critical radius $s_{\rm c}$ is around 1 $\mu$m, as the exponent $n$ decreases, the critical radius increases which means particles with larger radii are affected by the magnetic force.

Figure \ref{fig:2} shows the effect of the electrical charge polarity on the radial velocity of the dust grains. Solid curves correspond to a positively charged grain with one elementary charge and the dashed curves are for a negatively charged grain. In this plot, the magnetic exponent is $n=2$. Although for grains with radii larger than $s_{\rm c}$ the effect of magnetic field is negligible, for grains smaller than this critical size the direction of the radial velocity depends on the polarity of grain's charge. Outward mass flux is carried primarily by the positively charged grains with radii smaller than the critical radius $s_{\rm c}$. Note that this result is based on the assumed magnetic field configuration. We considered  a negative $\beta_{\theta}$ according to Lovelace, Romanova \& Newman (1994).

\begin{figure}
\plotone{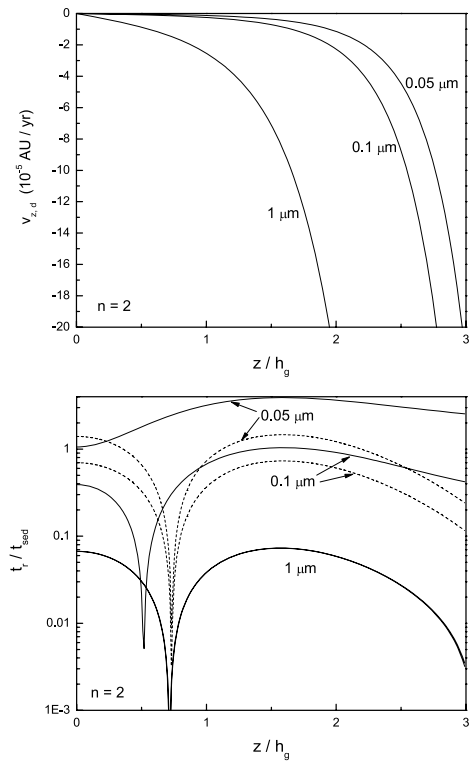}
\caption{({\it top}) Vertical velocity $v_{\rm z,d}$ of dust particles versus the distance from the midplane of the disc for grains with radii $s=0.05$ $\mu$m, $0.1$ $\mu$m and $1$ $\mu$m. The exponent of magnetic field is $n=2$ and charge of a particle is $q=-e$. ({\it bottom}) Ratio of the time scales $t_{\rm r}/t_{\rm sed}$ vs. distance from the midplane of the disc for the magnetic solutions (solid curves) and the nonmagnetic solutions (dashed curves).} \label{fig:3}
\end{figure}

Top plot of Figure \ref{fig:3} shows the vertical velocity $v_{\rm
z,d}$ of dust  particles vs. $z/h_{\rm g}$ for various sizes of
grains and $n=2$ and $q=-e$. Since the radial velocity significantly
increases for grains with $s<s_{\rm c}$ due to the magnetic force, we
expect larger vertical velocity comparing to the nonmagnetic solutions
according to equation (\ref{eq:vz1}). Although both the radial and
the vertical velocities increase due to the magnetic force for
grains with radii smaller than $s_{\rm c}$, sedimentation of the
particles can be determined based on the ratio of the time-scales of the sedimentation
and the radial migration.  The timescales of the sedimentation and
the radial migration are  defined as $t_{\rm sed}\approx z/|v_{\rm
z,d}|$ and  $t_{\rm r} \approx r/|v_{\rm r,d}|$, respectively.
Bottom plot of Figure \ref{fig:3} shows the ratio $t_{\rm r}/t_{\rm
sed}$  vs. distance from the midplane of the disc. Solid curves
correspond to the magnetic solutions and the dashed curves are for the
nonmagnetic solutions. Each curve is labeled by the radius of the grain.
As the size of grain decreases, the ratio of timescales increases.
When the effect of magnetic force becomes significant, the ratio of
the timescales is larger than the nonmagnetic case. For example, grains with
radius $0.05$ $\mu$m may sediment before any large migration in the
radial direction in the nonmagnetic case. But magnetic field force
significantly enhances the ratio of the timescales to values larger than
unity which implies large radial migration before sedimentation. Note that grains with radii 0.1, 1 and 10 $\mu$m have positive radial velocities near to the midplane (see Figure \ref{fig:1}). However, as the distance from the midplane increases, the radial velocity tends to zero at $z=z_0$. Beyond this location the radial velocity is negative. On the other hand, the radial migration timescale is proportional to inverse of the radial velocity. The dip at the curves occurs at the location where the radial velocity tends to zero (i.e. $z=z_0$). But for grains with radius $0.05$ $\mu$m, the radial velocity is negative for all the distances from the midplane. So, there is not a dip in its corresponding curve of the ratio of the timescales.

\section{Discussion and Summary}

We presented a model for the dynamics of charged grains in a magnetized, protoplanetary disc. Properties of the gas component correspond to a minimum-mass solar nebula and the toroidal and the poloidal components of the magnetic field are assumed to be power-law. The vertical and the radial velocities of the charged dust particles depend on the geometry of the magnetic field, charge and mass of the grains. We assumed that each particle is charged one elementary charge.

Our model illustrates the importance of charge of dust grains in their motions beyond the dead zone. Observations of protoplanetary discs, at least in the optical and infrared, mostly probe the upper disc layer. For charged grains  with radii smaller than $s_{\rm c}$,  effect of the magnetic force on their motion is stronger in the upper layers comparing to the midplane region. Since magnetic force prevents sedimentation of such grains to the midplane of the disc, one may expect grains with radii larger than a critical radius in midplane region, unless smaller grains are neutral. In a disc resembling to the minimum-mass solar nebula, and assuming that all grains are charged, our model predicts that grains in midplane of the disc are mostly larger than $s_{\rm c}=1$ $\mu$m. Dust sedimentation is known to affect the infrared spectra and the images of discs. In particular the far-infrared emission may be reduced by strong sedimentation. Although we did not investigate  spectra of the disc, it is interesting to study  possible modifications to the spectra due to the charge of the grains and the magnetic force. However, in a more realistic model not only the
coupling between the field lines and the charged grains should be
considered but also the significant effects of magnetic fields on
the gas component.

Takeuchi \& Lin (2003) studied the outflow of dust particles in the upper layers of optically disc. They showed that grains in surface layers are moving outward due to the stellar radiation. The outward mass flux is carried primarily by particles of size $0.1$ $\mu$m according to their calculations. Although we neglected the stellar radiation in our model, we expect strong outflows of sub-micron sized particles, if they are charged positively.  In model of Takeuchi \& Lin (2003), radiation pressure  is the main physical factor in making surface outflows of grains. But in our model, it is the magnetic force  which drives positively charged grains smaller than $s_{\rm c}$ to move outward. The actual optical depth along the radial direction in the midplane of the disk is huge, and thus, the radiation pressure from the central star onto the dust grains should be negligible. But neutral grains smaller than the critical radius in the midplane region can move outward once they charged positively. When they reach to the surface layers, not the magnetic force but the radiation pressure contribute to their outward motion. It will be interesting to study surface outflows under combined action of gas drag,  the radiation pressure, and the magnetic force in order to understand their relative importance in making surface outflows.

We can summarize our results:

(i) Motion of particles with radii smaller than $s_{\rm c}$ are affected by the magnetic field, but larger particles move independent of their charge and the magnetic force. For input parameters corresponding to a minimum-mass solar nebula, the critical radius is around 1 $\mu$m and so, sub-micron sized particles are affected by the field lines.

(ii) However, if the exponent of the magnetic field decreases or the charge of the grains increases, the critical radius $s_{\rm c}$ increases.

(iii) In our model, magnetic force increases the radial migration of charged particles comparing to when they are neutral. Also,  vertical velocity increases due to the magnetic force.

(iv) Although neutral dust grains tend to settle towards the midplane of the disc, magnetic force prevents charged grains with radii smaller than the critical radius from settling down.

(v) Direction of the radial motion is determined by the polarity of charged grains.

\acknowledgments

I am grateful to the anonymous referee for suggestions and comments to improve the paper. This research  was funded under the Programme for Research in
Third Level Institutions (PRTLI) administered by the Irish Higher
Education Authority under the National Development Plan and with partial
support from the European Regional Development Fund.

\end{document}